# Short Time and Structural Dynamics in Polypropylene Glycol Nanocomposite

M. Tyagi[1,2], R. Casalini[3], and C.M. Roland[3]

[1]NIST Center for Neutron Research, National Institute of Standards and Technology, 100 Bureau Drive, Gaithersburg, Maryland 20899-6102

[2]Department of Materials Science, University of Maryland, College Park, Maryland 20742

[3]Naval Research Laboratory, Chemistry Division, Washington DC 20375-5342

## ABSTRACT

The dynamics of polypropylene glycol, both neat and attached to silica nanoparticles, were investigated using elastic neutron backscattering and dielectric spectroscopy. The mean square displacement measured by the former is suppressed by the particles at temperatures corresponding to a dielectric secondary relaxation (that involves only a portion of the repeat unit) and the segmental relaxation (glass transition). Despite the suppression of the displacements, the motions are faster in the nanocomposite, primarily due to poorer packing (lower density) at the particle interface. At very low temperatures we discovered a new dynamic process in the polymer. Reflecting its very local nature, this process is unaffected by attachment of the chains to the silica.

## INTRODUCTION

The burgeoning interest in polymer nanocomposites is driven in large measure by the prospects for technologies that exploit their mechanical, electronic, optoelectronic, or magnetic properties. This potential can best be realized only with a detailed understanding of how particles having the dimensions of chain molecules can affect the configurations and motions of those molecules. Of course, a fundamental understanding of the influence of nano-reinforcement, confinement, and interfacial interaction is in itself a worthwhile objective, one that can offer new insights into the long-standing problem of the glass transition [1,2].

There is ample literature describing how nano-confinement and nano-particles affect macroscopic polymer properties, including the segmental dynamics in the vicinity of the glass transition [3,4,5,6]. A more recent discovery is that the dynamics at very short times, such as the vibrations and Boson peak [7,8,9,10] and the Johari-Goldstein (JG) secondary process [11,12] can be correlated with the structural relaxation ($\alpha$ process) occurring at long (*ca.* 100 s) times. The suggestion is that the glass transition is sensitive to motions occurring many decades earlier, implying that the high frequency dynamics may serve as a precursor to the transition. Additionally, there are processes that do not involve the entire molecule or repeat unit, such as pendant group motions, that also contribute at high frequencies [13,14,15,16,17]. However, their connection to structural relaxation is less apparent.

The fast motions observed in the mean squared displacement, $\langle u^2 \rangle$, probed by scattering experiments are often referred to as the caged dynamics, as the atoms rattle and eventually escape the local liquid structure defining this "cage". On warming in the glassy state, $\langle u^2 \rangle$ shows a marked increase in its temperature dependence at a characteristic temperature, $T^*$. The physics underlying this change at T* is of great interest, insofar as it bridges the local and mesoscopic dynamics. Cage dissolution has been associated with the onset of the JG process [18]. However, local, intramolecular reorientation of moieties with the molecule are also observed in the glassy state [13,14,15,16,17]. These are expected to transpire at higher frequency than the JG process.

Polypropylene glycol has attracted substantial research interest because its dielectric response is sensitive to both the segmental and chain motions. Its dielectric spectrum also exhibits two secondary relaxations at frequencies higher than the JG relaxation, a γ process [19,20] and as we report herein, a δ peak that we ascribe to methyl group rotation. The weak JG peak can only be resolved in measurements at high pressure [19,20,21] or upon extended physical aging [22]. These diverse motions of PPG, in combination with its propensity to form extended hydrogen bonding, results in rather complicated dynamics [23,24,25]. A prior study of PPG/silica nanocomposite [26] revealed that the relaxation times of polymer chains attached to the particles were shorter than for neat PPG.

In this work we address how nanoparticle reinforcement affects the mean square displacement measured by neutron scattering, and compare this to the effect on the dynamics measured by dielectric spectroscopy. The polymer and nanocomposite were those studied previously [26]. The nanocomposite was extracted with a poor solvent to remove unbound polymer, yielding a final silica content of 68% by volume. In this work we compare broadband dielectric measurements below the glass transition temperature, $T_g$, to elastic neutron backscattering to characterize the effect of silica particles on the dynamics at two length scales. We find that while the dielectric measurements show faster dynamics in the nanocomposite, the amplitude of the motions decreases, as reflected in $\langle u^2 \rangle$ measured by elastic neutron scattering. Reasons for this behavior are discussed.

EXPERIMENTAL

The PPG (Polysciences, Inc.) had a weight average molecular weight = 4.0 kg/mol; this corresponds to a chain coil size of about 5 nm. The Si nanoparticles (diameter = 12.5±2.5 nm) from Nissan Chemicals were functionalized with *ca.* 1,000 hydroxyl groups per particle. Isopropanol solutions/suspensions of these components were blended, sonicated, and then dried by heating in vacuo at 80°C. The sample was subsequently extracted with hexane at RT to remove unbound polymer. From thermogravimetric analysis, extraction reduced the polymer concentration from 78% to 32% by volume; thus, each silica particle has on average ~70 attached chains. This extracted sample was a compactable powder that lacked cohesive integrity, precluding mechanical measurements.

The High Flux Backscattering (HFBS) spectrometer at NIST was used in the elastic mode to characterize the high frequency dynamics of the polymer. The intensity of the elastically scattered neutrons was measured as a function of temperature over the range $0.25 \leq Q\ (\text{Å}^{-1}) \leq 1.75$, where Q is the scattering vector. The instrumental resolution was 0.8μeV, corresponding to 2 ns in the time domain.

Dielectric spectra were obtained with the sample between cylindrical electrodes (16 mm diameter) with a 0.1 mm separation. The permittivity was measured with a Novocontrol Alpha analyzer as a function to temperature, the latter controlled using a closed cycle helium cryostat (Cryo Ind.).

RESULTS

For harmonic oscillators in the long time limit, Gaussian behavior is obtained, with the mean squared displacement related to the neutron scattering intensity according to [27]

$$I(Q,T) = I_0(Q,T)\exp\left[-Q^2 \langle u^2(T) \rangle / 3\right] \qquad (1)$$

Although most atomic motions in soft condensed matter are anharmonic, this approximation is commonly used to characterize the dynamics of glass-forming materials. In the above equation, $I_0(Q,T)$ is the purely elastic intensity, usually measured on the sample at very low temperatures for which no dynamics is expected on the time scale of the backscattering spectrometer.

In Fig. 1 $\langle u^2 \rangle$ is shown for neat PPG and its nanocomposite in the temperature range from 4K to 300K. At very low temperatures, all atoms are essentially frozen on the time scale of the measurement, and therefore $\langle u^2 \rangle$ is relatively small, arising primarily from vibrations. As temperature increases, there is a linear increase in the mean square displacement with temperature, consistent with the assumption of harmonicity. The deviation from such dependency indicates additional dynamics appearing in the time window of the HFBS. The upturn in $\langle u^2 \rangle$ at $T^*$ around 100K is due to local motion. This process cannot be identified solely from the elastic scattering, and is due either to dissolution of the cage structure underlying the onset of the JG process and/or to unfreezing of local motions likely involving the chain end groups. The latter interpretation is consistent with previous dielectric measurements on PPG at elevated pressure [19,20,21]. The temperature associated with this change in dynamics is not markedly different for the neat and nanocomposite samples. However, the relaxation spectra show clearly a shift to higher frequency for chains bound to the nanoparticles (Figure 2).

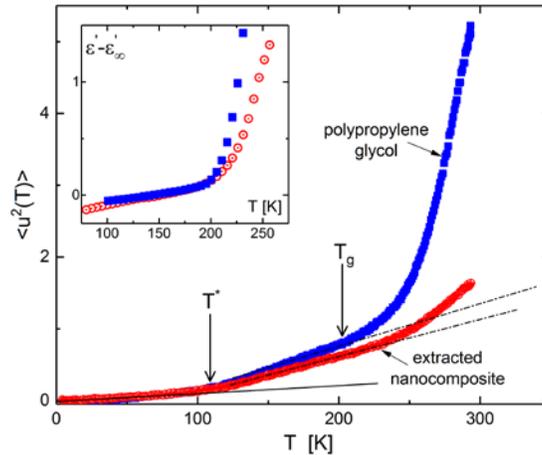

Figure 1. Mean squared displacement measured from elastic neutron scatting for neat PPG (squares) and the nanocomposite (circles). The lines are linear extrapolations, and vertical arrows indicate the approximate onset of secondary and structural dynamics. The inset show the dielectric permittivity after subtraction of the fit to the δ peak occurring at lower temperatures.

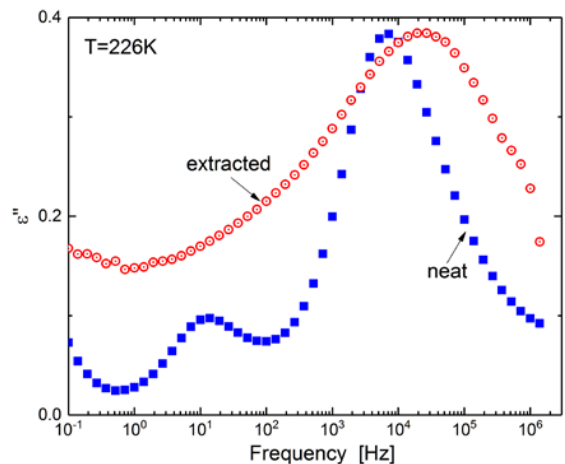

Figure 2. Dispersion in the dielectric loss at 226K showing the α peak and the normal mode (at lower frequency). Only for the neat material are the two peaks well resolved. For both there is a marked speeding up of the dynamics in the presence of the silica. The rise below 1 Hz is dc-conductivity due to mobile ions. The spectrum of the extracted polymer was multiplied by a constant to equalize the peak maxima.

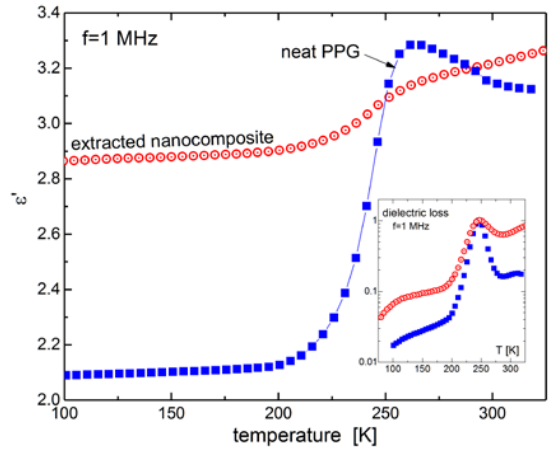

Figure 3. Permittivity for the polymer neat (squares) and nanocomposite (circles). The large response of the latter at low temperature is due to the silica. The inset shows the corresponding dielectric loss; the faster dynamics in the nanocomposite is more evident when the data are plotted as a function of frequency (see Fig. 2).

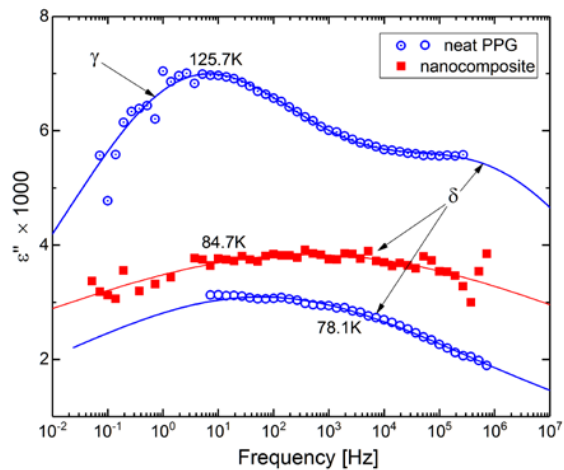

Figure 4. Dielectric spectra at low temperatures showing the weak $\delta$ process. The lower frequency $\gamma$ relaxation is evident in the highest temperature spectrum.

Above 100K the increase in $\langle u^2 \rangle$ is suppressed for the nanocomposite, a natural consequence of the tethering of the chains to the silica surface. The low temperature behavior is followed by a much stronger deviation from linearity around 200K due to structural relaxation (the glass transition). The increase of $\langle u^2 \rangle$ above 200K is strongly suppressed by the silica particles, congruent with the behavior of the secondary relaxation and tethering of the polymer chains to the silica particles. Similar behavior is exhibited by the iso-frequency dielectric constant measured as a function of temperature (Fig. 1 inset). This is expected since ε' is determined by the motion of permanent dipoles on the polymer chain; that is, their reorientation in the presence of an external electric field. The speeding up of the segmental dynamics (Fig. 2) is not evident in the change in either $\langle u^2 \rangle$ or ε' with temperature (Fig. 3). These measurements convolute temperature and rate dependences, obscuring changes in the peak frequency. To investigate further the change in segmental dynamics as seen in the dielectric results, dynamic structure factor measurements on these materials using the HFBS are planned.

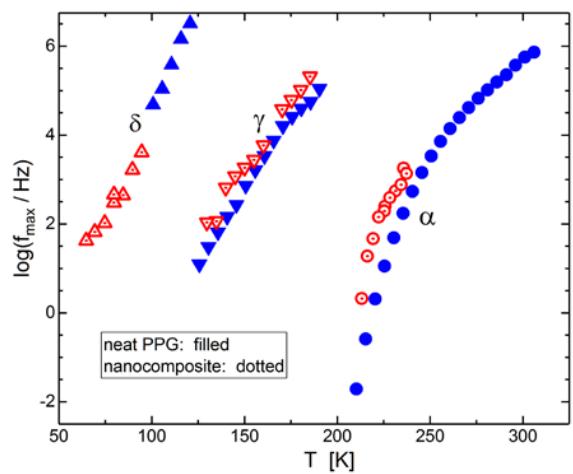

Figure 5. Peak frequencies for the segmental motion (circles), γ secondary relaxation (inverted triangles), and the δ process (triangles) for neat PPG (filled symbols) and the nanocomposite (dotted symbols). Note that only the weak δ peak observed at low temperatures is unaffected by the silica. The JG relaxation is subsumed by the α peak and cannot be resolved at ambient pressure.

Finally, we note that at very low temperatures (*ca.* 100K) there is a very weak dispersion (δ peak) in the dielectric spectrum, which has not been previously reported. The frequency of the underlying motion is too fast to be observed other than deeply in the glassy state. The obvious inference is that this dispersion is due to methyl group reorientation, since motion of this small pendant group is expected to be the fastest of the various dynamics in PPG.

## SUMMARY

The complex dynamics in PPG are illustrated in the relaxation map shown in Figure 5. All dynamics apart from the very local δ peak are faster in the extracted nanocomposite than in the neat polymer. (The β relaxation, previously shown to be a JG process [19,20,21], was not investigated herein because at ambient pressure it appears only as a high frequency wing on the more intense α dispersion. High pressure or physical aging is required to resolve the β peak.)

From the intensity of the γ peak in the dielectric spectrum, it can be assigned to reorientation of strongly polar groups in the chain molecule. An obvious possibility is non-bonded hydroxyl end groups [20], consistent with the increase in the γ peak frequency (relative to the frequency of the α peak) with PPG molecular weight [28] and with a negligible activation volume [19]. Since terminal hydroxyl moieties are expected to hydrogen bond to the silica surface, this attribution is consistent with the large decrease observed in ⟨$u^2$⟩ (Fig. 1).

It might be expected that slower motions would be the consequence of tethering chains to the nanoparticles; however, we have previously shown that packing within the interfacial region is poorer, and hence motions are faster [26]. However, as shown by the neutron scattering measurements, diffusion of the bound PPG chains is suppressed. Thus, nanoparticles can restrict the length scale of the dynamics, while accelerating the rate of motion. To further clarify this interesting, if not contradictory, dynamic behavior, we are presently measuring the full quasi-elastic spectra on these materials.

## ACKNOWLEDGEMENT

The work at NRL was supported by the Office of Naval Research. The HFBS at the NIST Center for Neutron Research is supported in part by the National Science Foundation under agreement number DMR-1508249. Certain commercial material suppliers are identified in this paper to foster understanding; such identification does not imply recommendation or endorsement by NIST.